\documentstyle[12pt,amssymbols]{article}
\newcommand{\bq}{\begin{equation}}
\newcommand{\eq}{\end{equation}}
\newcommand{\bqa}{\begin{eqnarray}}
\newcommand{\eqa}{\end{eqnarray}}
\newcommand{\ra}{\rightarrow}

\def\half{{1 \over 2}}

\def\D{\Delta}

\def\g{\gamma}

\def\ov{\over}

\def\rt{\sqrt{2}}
\def\ed{\end{document}}

\def\ra{\rightarrow}

\def\2pi{1\over 2\pi i}
\def\q{q-q^{-1}}
\def\newline{\hfil\break}

\def\ra{\rightarrow}
\def\va{\varphi}
\def\pa{\partial}
\def\sq2{\sqrt{2}}
\def\sqk2{\sqrt{2(k+2}}
\def\sqk{\sqrt{k}}

\def\be{\begin{equation}}
\def\ee{\end{equation}}
\def\br{\begin{array}}
\def\er{\end{array}}
\def\bea{\begin{eqnarray}}
\def\eea{\end{eqnarray}}
\def\ba{\begin{equation}\begin{array}}
\def\ea{\end{array}\end{equation}}
\def\bac{\begin{equation}\begin{array}{rll}}

\newcommand{\LS}{\widehat{sl(2)}}
\newcommand{\uq}{U_q (\widehat{sl(2)})}

\def\Z{{\Bbb Z}}

\begin{document}
\begin{titlepage}
\rightline{hep-th}
\rightline{CRM-2181}
\rightline{May 18, 1994}
\vbox{\vspace{15mm}}
\vspace{1.0truecm}
\begin{center}
{\LARGE \bf A Quantum Analogue of the Boson-Fermion
Correspondence}\\[8mm]
{\large A. Hamid Bougourzi$^{1}$  and  Luc Vinet$^{2}$}\\
[3mm]{\it Centre de Recherches Math\'ematiques,
Universit\'e de Montr\'eal\\
C.P. 6128-A, Montr\'eal (Qu\'ebec) H3C 3J7, Canada.}\\[30mm]
\end{center}
\begin{abstract}
We review the classical boson-fermion correspondence in the
 context of the $\widehat{sl(2)}$ current algebra at level 2.
This particular algebra is ideal to exhibit this
correspondence   because it can be realized either in terms
of three real bosonic fields or in terms of one real
and one complex fermionic fields. We also derive a fermionic
realization  of the quantum current algebra
$U_q(\widehat{sl(2)})$ at level 2 and by comparing
this realization with the existing bosonic one
we extend the classical correspondence to the
quantum case.
\end{abstract}
\footnotetext[1]{
Email: {\tt bougourz@ere.umontreal.ca}
}
\footnotetext[2]{
Email: {\tt vinet@ere.umontreal.ca}
}
\end{titlepage}
\section{Introduction}

It is now well established that the quantum affine
algebras play a key role in the description  of the
dynamics of
two-dimensional quantum integrable models such as
the $XXZ$ Heisenberg spin chain \cite{Daval92},
the sine-Gordon and the Thirring models \cite{Luk93}.

For  practical computational purposes however, these
infinite-dimensional symmetries are hardly directly
useful.
The reason is that the relevant representations are also
infinite-dimensional and their intertwiners or vertex operators,
which     are the main objects that explicitly appear in the computation
of physical quantities like correlation functions, form
 factors and S matrix elements, have nontrivial fusion rules
(operator product expansions.) This makes
their normal ordering highly complicated and very difficult to
use for the computation  of the above physical quantities.
There are no  simple analogues here of the Wick theorem which
allows to perform the normal ordering of
operators constructed in terms of free fields.
However, one can still hope to apply the Wick theorem to
vertex operators if one succeeds to  realize these
intertwiners in terms
of free fields.
 This has been the main   motivation for obtaining
the free field realizations of all the infinite-dimensional
symmetry
algebras that appear in conformal field theories and more
recently, for the extension of these realizations to
quantum affine algebras. (See Ref. \cite{DiFr93} for a general
review, more references will be given in the
subsequent sections of this paper.)

In contrast  to the situation with the classical affine algebras,
so far, all  the existing free field realizations of
quantum affine algebras have only been given  in terms
of bosonic fields. There is a single
exception to this statement:
a fermionic construction of these algebras at level zero
has been derived in Ref. \cite{Hay90}, it does not
seem possible however to  generalize this
construction  to higher levels.
Since quantum affine algebras with positive levels  are
these that play a major role in physical models through their
highest weight representations, it is therefore of interest
to investigate fermionic realizations of these algebras
 so as to provide  alternative models to the existing   bosonic ones.

In this paper, we construct a fermionic realization
for  $\uq$  at the particular level 2.
The main feature of this algebra is that its
fermionic realization requires one real
and one complex fermionic fields. Since we know already
the bosonization of this algebra, we may therefore derive
the equivalence between the two and obtain a quantum
analogue of the important boson-fermion correspondence that plays
a major role in conformal field theories \cite{Fre81,
GoOl86}.

This paper is organized as follows. In section 2, we first
review the classical $\LS$ algebra at arbitrary level in order
to  establish our notations. Then,
we  describe both  bosonic and fermionic realizations of this
algebra at the particular level 2 and
discuss  their equivalence, i.e.,
the classical boson-fermion correspondence. We give in section 3
the quantum analogues of the results of section 2.
We first define the $\uq$ quantum algebra at
arbitrary level and describe one of its
bosonizations
at level 2. Then, we present the main new results of this
paper, that is, the fermionic realization of  $\uq$
 at level 2 and its equivalence to the
bosonic one, i.e., the quantum analogue of the boson-fermion
correspondence. Conclusions are found in section 4.
\section{Classical boson-fermion correspondence}

\subsection{The $\LS$ current algebra}

The $\LS$ affine algebra is generated by the
Chevalley basis elements
$\{e_i, f_i, h_i, d; i=0,1\}$ subjected to
\bac
{[h_i, h_j]}&=&0,\\
{[h_i,e_j]}&=&a_{ij}e_j,\\
{[h_i,f_j]}&=&-a_{ij}f_j,\\
{[e_i,f_j]}&=&\delta_{ij}h_i,\\
{[d,h_i]}&=&0,\\
{[d,e_i]}&=&\delta_{i0}e_i,\\
{[d,f_i]}&=&-\delta_{i0}f_i,\\
{[k,x]}&=&0,\quad \forall x\in \LS,\quad i,j=0,1;
\label{def}
\ea
where,
\be
(a_{ij})=\pmatrix{2&-2\cr -2&2}
\label{Cartan}
\ee
is the $\LS$ generalized Cartan matrix and  the element
$k=h_1+h_0$  referred to as the level.
The Serre relations should be added to the defining relations
(\ref{def}).
As is well     known, the above algebra is isomorphic
to the  one generated by the elements
$\{E^{\pm}_n,H_n,k,d;n\in\Z\}$ which obey:
\bac
{[H_n,H_m]}& = & nk\delta_{n+m,0},\\
{[H_n,E^{\pm}_m]}&=&
\pm\sqrt{2}E^\pm_{n+m},\\
{[E^+_n,E^-_m]} &=& \sqrt{2}H_{n+m}+nk\delta_{n+m,0},\\
{[d,E^{\pm}_n]}&=&nE^{\pm}_n,\\
{[d,H_n]}&=&nH_n,\\
{[k,x]}&=&0,\quad \forall x\in \LS;\quad n,m\in\Z.
\label{da}
\ea
The  isomorphism is realized through the
identifications
\bac
h_1&\equiv& \sq2 H_0,\quad h_0\equiv k-\sq2 H_0,\\
e_1&\equiv& E^+_0,\quad e_0\equiv E^-_1,\\
f_1&\equiv& E^-_0,\quad f_0\equiv E^+_{-1}.
\ea
Note that the elements $k$ and $d$ are not redefined in this
second form of $\LS$. There is yet a third form
of this algebra which is isomorphic to the previous two
and which we refer to as the $\LS$ current
algebra. It is most useful for the
purpose of discussing
free field realizations as will be clear
throughout the remainder of this paper.
The structure relations are given as operator
product expansions (OPE's) among the currents
(i.e., formal generating functions
in a complex  variable
$z$) $E^\pm(z)=\sum_{n\in\Z}E^\pm_nz^{-n-1}$ and
$H(z)=\sum_{n\in\Z}H_nz^{-n-1}$. They read
\bea
H(z).H(w)&\sim &{k\over (z-w)^2},\quad |z|>|w|,
\label{ope1}\\
H(z).E^{\pm}(w)&\sim& \pm{\sqrt{2}E^\pm(w)\over z-w},
\quad |z|>|w|,
\label{ope2}\\
E^+(z).E^-(w)&\sim &{\sqrt{2}H(w)\over z-w}+{k\over (z-w)^2},
\quad |z|>|w|,      \label{ope3}\\
E^-(z).E^+(w)&\sim &{\sqrt{2}H(w)\over z-w}+{k\over (z-w)^2},
\quad |z|>|w|.      \label{ope4}
\eea
Here the symbol $\sim$  means  an equality up to
regular terms as $z$ approaches $w$. The relations involving
both elements $k$ and $d$ are the same as in (\ref{da}).

\subsection{Bosonic realization of $\LS$ at level 2}

We now briefly review the bosonization of the
$\LS$ current algebra given in
(\ref{ope1})-(\ref{ope4}). This bosonization    is  called the
Wakimoto realization and is
valid for a generic level $k$.
Here, we shall only present the case $k=2$ which is in fact very
similar to the general one. It requires three free
real bosonic fields
$\phi^1(z)$, $\phi^2(z)$ and $\phi^3(z)$ with OPE's
\be
\phi^i(z).\phi^j(w)=-(-1)^{i-1}\delta^{ij}\ln(z-w)+
:\phi^i(z)\phi^j(w):,\quad |z|>|w|.
\label{hca}
\ee
The mode expansions of these fields are given by
\be
\phi^j(z)=\phi^j-i\phi^j_0\ln{z}+i\sum_{n\neq 0}
{\phi^j_n\over  n}
z^{-n},\qquad j=1,2,3.
\label{cphi}
\ee
The bosonic modes $\{\phi^j_n, \phi^j; n\in \Z; j=1,2,3\}$
generate a
 Heisenberg algebra with non-vanishing commutators given by
\bac
{[\phi^j_n,\phi^\ell_m]}&=&(-1)^{j-1}n\delta^{j\ell}
\delta_{n+m,0},\\
{[\phi^j,\phi^\ell_0]}&=&(-1)^{j-1}i\delta^{j\ell},
\qquad\qquad  j=1,2,3.
\label{hma}
\ea
In (\ref{hca}), the symbol :: denotes the bosonic normal
ordering which is defined by
\bac
: \phi^1_n\phi^1_m:&=& \phi^1_m\phi^1_n, \qquad n>0,\\
 : \phi^1_n\phi^1_m:&=& \phi^1_n\phi^1_m, \qquad n<0,\\
: \phi^1_0\phi^1:&=& \phi^1\phi^1_0,\\
: \phi^1\phi^1_0:&=& \phi^1\phi^1_0.
\label{nor}
\ea
This means that the creation modes $\{\phi^i_n; n<0\}$ and
the shift mode $\phi^i$ are always moved to the left of the
annihilation modes $\{\phi^i_n; n>0\}$ and of the momentum  mode
$\phi^i_0$ in products of the above bosonic fields.
With these definitions it can easily be checked that
the `Heisenberg mode algebra' (\ref{hma}) is equivalent
to the `Heisenberg current algebra' (\ref{hca}).
As usual in the classical case (i.e. $q=1$), a field  or
a product
of fields defined at the same point $z$ is understood to be
normal ordered and the symbol ::
is omitted.
The Wakimoto bosonization reads
\bac
H(z)&=&i\sq2\pa\phi^1(z),
\label{H}\\
E^\pm(z)&=&\left(\pm i\pa\phi^2(z)+
i\sq2
\pa\phi^3(z)\right)\exp\{\pm  i(\phi^1(z)+\phi^2(z))\}.
\label{br}
\ea

\subsection{Fermionic realization of $\LS$ at level 2}

Unlike the Wakimoto bosonization, the fermionic realization
of the $\LS$ current algebra (\ref{ope1})-(\ref{ope4})
exists only for the particular cases of $k=0,1,2.$
Let us review it when $k=2$. It requires one real and
one complex  fermionic fields, which we denote
respectively by $\psi(z)$ and   $\chi(z)$. These fields
satisfy the following OPE's:
\bac
\psi(z).\psi(w)&=&{1\over z-w}+:\psi(z)\psi(w):,\quad |z|>|w|,\\
\chi(z).\chi^{\dag}(w)&=&{1\over z-w}+:\chi(z)\chi^{\dag}(w):,
\quad |z|>|w|,\\
\chi^{\dag}(z).\chi(w)&=&{1\over z-w}+:\chi^{\dag}(z)\chi(w):,
\quad |z|>|w|,\\
\chi(z).\chi(w)&=&:\chi(z)\chi(w):,\\
\chi^{\dag}(z).\chi^{\dag}(w)&=&:\chi^{\dag}(z)\chi^{\dag}(w):,
\label{cca}
\ea
where $\chi^{\dag}(z)$ stands for the Hermitian conjugate
of         $\chi(z)$.
The mode expansions of these fields are given by
\bac
\psi(z)&=&\sum_{r\in \Z+1/2}\psi_r z^{-r-1/2},\\
\chi(z)&=&\sum_{r\in \Z+1/2}\chi_r z^{-r-1/2},\\
\chi^{\dag}(z)&=&\sum_{r\in \Z+1/2}\chi^{\dag}_r z^{-r-1/2}.
\ea
The fermionic modes $\{\psi_r, \chi_r, \chi^{\dag}_r; r\in \Z+1/2\}$
satisfy the following anticommutation relations:
\bac
\{\psi_r,\psi_s\}&=&\delta_{r+s,0},\\
\{\chi_r,\chi^{\dag}_s\}&=&\delta_{r+s,0},\\
\{\chi_r,\chi_s\}&=&0,\\
\{\chi^{\dag}_r,\chi^{\dag}_s\}&=&0,\quad r,s\in\Z+1/2;
\label{ccma}
\ea
with $\psi_r$ commuting with
$\{\chi_r, \chi^{\dag}_r\}$.
Note that the above `Clifford mode algebra'
(\ref{ccma}) is equivalent to
the `Clifford current algebra' (\ref{cca}) if the fermionic
normal ordering introduced in (\ref{cca}) is defined
as in the bosonic case,
that is, in products of the
fermionic fields, the creation modes
$\{\psi_r, \chi_r,
\chi^{\dag}_r; r\leq -1/2 \}$ are always moved to the left
of the annihilation modes $\{\psi_r, \chi_r,
\chi^{\dag}_r; r\geq 1/2\}$.
Here also we adopt the standard  convention
that a field  or a product
of fields defined at the same point $z$ is understood to be
normal ordered, in which case the symbol :: will be omitted.
It can easily be checked that with this normal ordering
the following relation is satisfied:
\be
:\psi(z)\psi(z):=0.
\label{pp}
\ee
With the above definitions and conventions, the fermionic
realization of the $\LS$ current algebra at level 2
(\ref{ope1})-(\ref{ope4}) is given by
\bac
H(z)&=&\sq2\chi(z)\chi^{\dag}(z),\\
E^+(z)&=&\sq2\psi(z)\chi(z),\\
E^-(z)&=&\sq2\psi(z)\chi^{\dag}(z).
\label{fr}
\ea
Let us show for example how (\ref{ope3}) is simply
recovered from these identifications:
\bac
E^+(z).E^-(w)&=&2\psi(z)\chi(z).\psi(w)\chi^{\dag}(w)\\
&=&2\left({1\over z-w}+:\psi(z)\psi(w):\right)
\left({1\over z-w}+:\chi(z)\chi^{\dag}(w):\right)\\
&=&{2\over (z-w)^2}+
{2:\chi(w)\chi^{\dag}(w):\over z-w}
+{\rm regular\> as}\>\> z\ra w\\
&=&{2\over (z-w)^2}+
{\sq2 H(w)\over z-w}
+{\rm regular\> as}\>\> z\ra w.
\label{cla}
\ea
Note that relation (\ref{pp}) has been used in this
derivation.

\subsection{Classical boson-fermion correspondence}

In this section, the bosonic and
the fermionic realizations are  seen to be
equivalent through
identifications among the fermionic fields and
certain vertex operators constructed in terms of
the bosonic free fields. This is known as
the boson-fermion correspondence.
We prove to start that the complex fermionic
field $\chi(z)$, its conjugate $\chi^{\dag}(z)$ and
their quadratic normal ordered products can be
bosonized as follows:
\bac
\chi(z)&\equiv& e^{i\phi^1(z)},\\
\chi^{\dag}(z)&\equiv& e^{-i\phi^1(z)},\\
:\chi(z)\chi^{\dag}(w):&\equiv& {1\over z-w}
\left(:e^{i\phi^1(z)-i\phi^1(w)}:-1\right),
\label{chch}\\
:\chi^{\dag}(z)\chi(w):&\equiv& {1\over z-w}
\left(:e^{-i\phi^1(z)+i\phi^1(w)}:-1\right),\\
:\chi(z)\chi(w):&\equiv& (z-w)
:e^{i\phi^1(z)+i\phi^1(w)}:,\\
:\chi^{\dag}(z)\chi^{\dag}(w):&\equiv& (z-w)
:e^{-i\phi^1(z)-i\phi^1(w)}:.
\label{cfr}
\ea
Indeed, using
the relation (which describes the normal ordering
of pure exponential  vertex operators):
\be
e^{ia\phi^1(z)}.e^{ib\phi^1(w)}=(z-w)^{ab}
:e^{ia\phi^1(z)+ib\phi^1(w)}:,\quad a,b\in R,
\ee
one can easily show that the
current algebra generated by the vertex operators
$\exp{i\phi^1(z)}$ and $\exp{-i\phi^1(z)}$ is
the same as the one generated by the fermionic fields
$\chi(z)$ and $\chi^{\dag}(z)$ (\ref{cca}).
More explicitly, we have
\bac
e^{i\phi^1(z)}.e^{-i\phi^1(w)}&=&
{1\over z-w}+{1\over z-w}
\left(:e^{i\phi^1(z)-i\phi^1(w)}:-1\right),\\
e^{-i\phi^1(z)}.e^{i\phi^1(w)}&=&
{1\over z-w}+{1\over z-w}
\left(:e^{-i\phi^1(z)+i\phi^1(w)}:-1\right),\\
e^{i\phi^1(z)}.e^{i\phi^1(w)}&=&
(z-w)
:e^{i\phi^1(z)+i\phi^1(w)}:,\\
e^{-i\phi^1(z)}.e^{-i\phi^1(w)}&=&
(z-w)
:e^{-i\phi^1(z)-i\phi^1(w)}:.
\ea
Note that if we expand  both sides of the relation
(\ref{chch}) around $z=w$ we get at the zeroth order
the relation
\be
:\chi(w)\chi^{\dag}(w):=i\pa\phi^1(w).
\ee
Re-written in terms of modes, this relation leads to
\bac
\phi^1_n&=&\sum_{r\in \Z+1/2}:\chi_r\chi^{\dag}_{n-r}:\\
&=&\sum_{r\leq -1/2}\chi_r\chi^{\dag}_{n-r}-
\sum_{r\geq 1/2}\chi^{\dag}_{n-r}\chi_r,\quad n\in \Z.
\label{bla}
\ea
Therefore a Heisenberg mode algebra generated by
$\phi^1_n$ can always be realized in terms of a
Clifford algebra generated by  $\chi_r$ and
$\chi^{\dag}_r$. This explains the correspondence between
a free real bosonic field and a free complex fermionic field.

Let us make here the
following remark: since $\chi^{\dag}(z)$ is the Hermitian
conjugate  of $\chi(z)$ and because of the above
identifications, we expect
 the vertex operator $e^{-i\phi^1(z)}$
to be also the Hermitian conjugate of $e^{i\phi^1(z)}$.
To check  this, recall the general
definition of the adjoint operator of a vertex operator.
Let
\bac
U_{\alpha}(z)&=&:e^{i\alpha \phi^1(z)}:\\
&\equiv&e^{i\alpha \phi}e^{-\alpha\sum_{n<0}
{\phi_n\over n}z^{-n}}  e^{-\alpha\sum_{n>0}
{\phi_n\over n}z^{-n}}e^{\alpha \phi_0\ln z}
\label{HC1}
\ea
be a vertex operator with a `charge'
$\alpha$ ( in our case $\alpha=1$.)
Due to  nontrivial scaling dimension
(conformal dimension) of $U_{\alpha}(z)$,
its Hermitian conjugate is
defined by \cite{Gin88,GoOl86}
\be
[U_\alpha(z)]^{\dag}=U_{-\alpha}({1\over z^*})
{1\over {z^*}^{\alpha^2}}.
\label{HC2}
\ee
In analogy with
classical functions, we might think of $U_{-\alpha}
(z)$ as the Hermitian conjugate of $U_{\alpha}(z)$
provided that $\phi^1(z)$ is real, that is,
\be
[\phi^1(z)]^{\dag}=\phi^{1}({1\over z^*}),
\label{HC3}
\ee
since the scaling dimension of $\phi^1(z)$ is equal
to zero. The latter relation reads in terms
of modes as
\bac
\phi^{1\dag}&=&\phi^1,\\
\phi^{1\dag}_0&=&\phi^1_0,\\
\phi^{1\dag}_n&=&\phi^1_{-n}.
\label{adj}
\ea
With these definitions of the adjoint modes $\{
\phi^{1\dag},\phi^{1\dag}_n;n\in \Z\}$, we can easily
verify  that $U_{-\alpha}
(z)$ is the Hermitian conjugate of $U_{\alpha}(z)$,
i.e.,   that the relation (\ref{HC2}) is satisfied. Indeed,
\bac
[U_\alpha(z)]^{\dag}&=&
[e^{i\alpha \phi^1}e^{-\alpha\sum_{n<0}
{\phi^1_n\over n}z^{-n}}  e^{-\alpha\sum_{n>0}
{\phi^1_n\over n}z^{-n}}e^{\alpha \phi^1_0\ln z}]^{\dag}\\
&=&
e^{\alpha \phi^1_0\ln z^*}e^{-i\alpha \phi^1}
e^{-\alpha\sum_{n>0}
{\phi^1_{-n}\over n}z^{-n*}}  e^{-\alpha\sum_{n<0}
{\phi^1_{-n}\over n}z^{-n*}}\\
&=&z^{*-\alpha^2}
e^{-i\alpha \phi^1}e^{\alpha\sum_{n<0}
{\phi^1_n\over n}z^{*n}}  e^{\alpha\sum_{n>0}
{\phi^1_n\over n}z^{*n}}e^{-\alpha \phi^1_0\ln z^{-1*}}\\
&=&U_{-\alpha}({1\over z^*})
{1\over {z^*}^{\alpha^2}}.
\ea
For consistency,  the  adjoint operation
as defined in (\ref{adj})
must be an antiautomorphism of
the Heisenberg algebra (\ref{hma});
it is readily checked that this is  the case.

Let us now describe the bosonization of the real
fermion $\psi(z)$. This case is more subtle than the
previous one because if we compare both the bosonic and
fermionic realizations of $\LS$ as given respectively
by (\ref{br}) and (\ref{fr}) and if we take into account
the bosonization of the complex fermions $\chi(z)$ and
$\chi^{\dag}(z)$ (\ref{cfr}), it is then clear that there are
two different ways of bosonizing the real fermionic
field $\psi(z)$. They are given by
\be
\psi(z)\equiv \psi^\pm(z)
=\left(\pm {i\over \sq2}\pa\phi^2(z)+
i\pa\phi^3(z)\right)e^{\pm i\phi^2(z)},
\label{psps}
\ee
where we have introduced the notation
$\psi^\pm(z)$ in order to distinguish these two possible
bosonizations of $\psi(z)$ from each other.
This peculiar feature-the
existence of two possible bosonizations of the
same real fermion $\psi(z)$-deserves some explanation.
Given the form of the vertex
operators arising in  the bosonizations
of $\psi^\pm(z)$ and the fact that the OPE's of the fields $\phi^2(z)$
and $\phi^3(z)$ involve a `Lorentzian metric,' that is,
\bac
\phi^2(z).\phi^2(w)&=&\ln(z-w)+:\phi^2(z)\phi^2(w):,\\
\phi^3(z).\phi^3(w)&=&-\ln(z-w)+:\phi^3(z)\phi^3(w):,
\ea
we need to derive a bosonic normal ordering between
two generic vertex operators of the form
$V(z,a,a^\prime)=ia.\pa\phi(z)\exp{ia^\prime.\phi(z)}$
and $V(z,b,b^\prime)=ib.\pa\phi(z)\exp{ib^\prime.\phi(z)}$.
Here $\phi(z)\equiv(\phi^2;\phi^3)$. The symbols $a$,
$a^\prime$, $b$ and $b^\prime$ stand for
two-dimensional vectors with real number entries and
their
scalar product is defined
through the metric $diag\>(-;+)$. This normal
ordering is given by
\bac
V(z,a,a^\prime).V(w,b,b^\prime)
=&&
:\{{a.b-(a.b^\prime)(a^\prime.b)\over (z-w)^2}+
{i(a.b^\prime)b.\pa\phi(w)-i(a^\prime.b)a.\pa\phi(z)\over
z-w}\\
&& +(ia.\pa\phi(z))(ib.\pa\phi(w))\}
 (z-w)^{a^\prime.
b^\prime}
e^{ia^\prime.\phi(z)+ib^\prime.\phi(w)}:.
\label{vono}
\ea
$\psi^+(z)$ and $\psi^-(z)$ can be respectively
identified with $V(z,a,a^\prime)$ and $V(z,b,b^\prime)$ if we
set
\bac
a&=&(-{1\over \sq2};1),\\
a^\prime&=&(-1;0),\\
b&=&({1\over \sq2};1),\\
b^\prime&=&(1;0).
\ea
Using these identifications and the vertex operator
normal ordering (\ref{vono}) we get the following OPE's:
\bac
\psi^+(z).\psi^-(w)&\equiv&
V(z,a,a^\prime).V(w,b,b^\prime)\\
&=&{1\over z-w}+{\rm regular\> as}\>\> z\ra w,\\
\psi^-(z).\psi^+(w)&\equiv&
V(z,b,b^\prime).V(w,a,a^\prime)\\
&=&{1\over z-w}+{\rm regular\> as}\>\> z\ra w,\\
\psi^+(z).\psi^+(w)&\equiv&
V(z,a,a^\prime).V(w,a,a^\prime)\\
&=&{I^+(w)\over z-w}+{\rm regular\> as}\>\> z\ra w,\\
\psi^-(z).\psi^-(w)&\equiv&
V(z,b,b^\prime).V(w,b,b^\prime)\\
&=&{I^-(w)\over z-w}+{\rm regular\> as}\>\> z\ra w,
\ea
where the fields $I^{\pm}(w)$  are given by
\bac
I^+(w)&=&:\left((ia.\pa\phi(w))^2+{i\over \sq2}a.\pa^2\phi(w)
\right)e^{2ia^\prime.\phi(w)}:,\\
I^-(w)&=&:\left((ib.\pa\phi(w))^2+{i\over \sq2}b.\pa^2\phi(w)
\right)e^{2ib^\prime.\phi(w)}:.
\ea
Therefore, the real fermionic field $\psi(z)$ can be bosonized
in two different ways if the following conditions are
satisfied:
\be
I^{\pm}(w)\equiv 1.
\label{cid}
\ee
This is indeed the case and is a general feature of the
bosonizations of conformal field theories with background
charges \cite{Bil89}. The background charge
can here  be easily read off from
the bosonization of the energy-momentum tensor $T(z)$, which is in turn
obtained from the bosonization  of the currents $H(z)$ and $E^\pm(z)$
through the Sugawara construction. It reads
\be
T(z)=-1/2(\pa\phi^1(z))^2-1/2(\pa\phi(z))^2+i\alpha_0
.\pa^2\phi(z),
\ee
with the two dimensional background charge $\alpha_0$
given by
\be
\alpha_0=(0;-{1\over 2\sq2}).
\ee
In fact, with this background charge, we can already check
that the fields $I^{\pm}(w)$  have the same
conformal dimension as the identity field 1. Let us recall
for this purpose that the conformal dimension $\Delta$ of a vertex
field of the form $ix.\pa^{n}\phi(w)\exp{iy.\phi(w)}$ or
$(ix.\pa\phi(w))^{n}\exp{iy.\phi(w)}$, with $x$ and
$y$ arbitrary two-dimensional
vectors, is given by
\be
\Delta=n+{y^2\over 2}-\alpha_0.y.
\ee
Applying this formula to  $I^+(w)$ and $I^-(w)$
we find that both have  zero conformal dimension just as
the identity field.

To sum up this analysis of the classical case:
the $\LS$ current algebra at
level 2 admits both
bosonic and fermionic realizations, which are
described respectively by the relations
(\ref{H}) and (\ref{fr}); the equivalence between
the two, the so-called boson-fermion correspondence,
is given by the relations (\ref{cfr}) and (\ref{psps}).
\section{Quantum boson-fermion correspondence}
\subsection { The $\uq$ quantum current algebra}

The $\uq$ quantum algebra can be presented as
follows in terms of the Chevalley basis elements
$\{e_i,f_i,t^{\pm 1}_i,q^{\pm d};  i=0,1\}$
\cite{Dri85,Jim85,Dri86}. The defining relations
are taken to be
\pagebreak
\bac
&&t_it_j=t_jt_i,\\
&&t_i e_j t^{-1}_i=q^{a_{ij}}e_j,\\
&&t_i f_j t^{-1}_i=q^{-a_{ij}}f_j,\\
&&[e_i, f_j]=\delta_{ij}{t_i-t_i^{-1}\over q-q^{-1}},\\
&&q^de_iq^{-d}=q^{\delta_{i0}}e_i,\\
&&q^df_iq^{-d}=q^{-\delta_{i0}}f_i,\\
&&q^dt_iq^{-d}=t_i,
\label{cheval}
\ea
with $(a_{ij})$ the $\LS$ generalized Cartan
matrix,
and supplemented with quantum
analogues of the Serre relations.
This algebra is an associative Hopf algebra with
comultiplication
\bac
\Delta(e_i)&=&e_i\otimes 1+t_i\otimes e_i,\\
\Delta(f_i)&=&f_i\otimes t_i^{-1} +1\otimes f_i, \\
\Delta(t^{\pm 1}_i)&=&t^{\pm 1}_i\otimes t^{\pm 1}_i,\\
\Delta(q^{\pm d})&=&q^{\pm d}\otimes q^{\pm d},
\label{comul}
\ea
and antipode
\bac
S(e_i)&=&-t^{-1}_ie_i,\\
S(f_i)&=&-f_it_i,\\
S(t^{\pm 1}_i)&=&t_i^{\mp 1},\\
S(q^{\pm d})&=&q^{\mp d}.
\label{antip}
\ea
There is a second equivalent definition of $\uq$,
which was first given by Drinfeld \cite{Dri88}. It is generated
by the elements
$\{ E^{\pm}_n,H_m ,q^{\pm \rt H_0},
q^{\pm d}, \g^{\pm 1/2}; n\in \Z,~m \in
\Z_{ \neq 0}\}$ that obey
\bac
&&{[H_n,H_m]} = {[2n]\over 2n} {{\g^{nk}-\g^{-nk}}
\ov {q-q^{-1}}}
\delta_{n+m,0},\qquad
n\neq 0,\\
&&{[q^{\pm\rt H_0}, H_m]}=  0,\\ && {[H_n,E^{\pm}_m]}=
\pm\sqrt{2}{\g^{\mp |n|k/2}[2n]\over 2n}
E^\pm_{n+m}, \qquad n\neq 0,\\
&&  q^{\rt H_0}  E^\pm_n q^{-\rt H_0} =q^{\pm 2} E^\pm_n,\\
&&{[E^+_n,E^-_m]} = {\g^{k(n-m)/2}\Psi_{n+m}-\g^{k(m-n)/2}
\Phi_{n+m}\over q-q^{-1}},\\
&&E^\pm_{n+1}E^\pm_m-q^{\pm 2}E^\pm_mE^\pm_{n+1}=
q^{\pm 2}E^\pm
_nE^\pm_{m+1}-E^\pm_{m+1}E^\pm_n,\\
&&q^dE^\pm_nq^{-d}=q^nE^\pm_n,\\
&&q^dH_nq^{-d}=q^nH_n,\\
&&{[\g^{\pm 1/2}, x]}=0,\quad \forall x\in \uq.
\label{cwb}
\ea
We are using the  notation
$[n]\equiv (q^n-q^{-n})/(q-q^{-1})$, $k$ is again called the
level, and, $\Psi_n$ and $\Phi_n$
are given by
the mode expansions of the fields
$\Psi(z)$ and $\Phi(z)$, which are themselves defined by
\bac
\Psi(z)&=&\sum\limits_{n\geq 0}\Psi_nz^{-n}=q^{\sqrt{2}H_0}
\exp\{\sqrt{2}(\q)\sum\limits_{n>0}H_nz^{-n}\},\\
\Phi(z)&=&\sum\limits_{n\leq 0}\Phi_nz^{-n}=q^{-\sqrt{2}H_0}
\exp\{-\sqrt{2}(\q)\sum\limits_{n<0}H_nz^{-n}\}.
\label{algebra}
\ea
The isomorphism between these two presentations of
$\uq$  is provided by the following identifications:
\bac
t_0&\equiv&\gamma^k q^{-\rt H_0},\\
t_1&\equiv&q^{\rt H_0},\\
e_0&\equiv&E^-_1q^{-\rt H_0},\\
e_1&\equiv&E^+_0,\\
f_0&\equiv&q^{\rt H_0}E^+_{-1},\\
f_1&\equiv&E^-_0.
\label{ident}
\ea
Using these  assignments, the
comultiplication (\ref{comul}) can be rewritten
as follows in terms of the Drinfeld generators:
\be\br{rcl}
\!\!\!\!\!\!& &\Delta(E^+_n)=E^+_n\otimes\gamma^{kn}+
\gamma^{2kn}q^{\sq2 H_0}\otimes
E^+_n+ \sum_{i=0}^{n-1}\gamma^{k(n+3i)/2}
\Psi_{n-i}\otimes \gamma^{k(n-i)}
 E^+_i\: {\rm mod}   \: {N_-}\otimes {N_+^2},\\
 \!\!\!\!\!\!& &\Delta(E^+_{-m})=
E^+_{-m}\!\otimes\!\gamma^{-km}\!+\!
q^{-\sq2 H_0}\!\otimes\! E^+_{-m}+
\sum_{i=0}^{m-1}\gamma^{{k(m-i)\over 2}}
\Phi_{-m+i}\otimes \gamma^{k(i-m)}
 E^+_{-i}
\: {\rm mod  }\:
 N_-\otimes N_+^2,\\
\!\!\!\!\!\!& &\Delta(E^-_{-n})=
E^-_{-n}\!\otimes\!\gamma^{-2kn}q^{-\sq2
H_0}\!+\!
  \gamma^{-kn}\!\otimes \!E^-_{-n}\!+\!\!
\sum_{i=0}^{n-1}\!\gamma^{-k(n-i)}
  E^-_i\!\otimes\!\gamma^{{-k(n+3i)\over 2}}
\Phi_{i-n}\: {\rm mod  }\:
 N_-^2\!\otimes\! N_+,\\
\!\!\!\!\!\!& &\Delta(E^-_m)=\gamma^{km}
\otimes E^-_m+E^-_m\otimes
q^{\sq2 H_0} +
\sum_{i=1}^{m-1}\gamma^{k(m-1)}E^-_m
\otimes \gamma^{-k(m-i)/2}
\psi_{m-i}\:{\rm mod  }\: N_-^2\otimes N_+,\\
\!\!\!\!\!\!& &\Delta(H_m)=H_m\otimes\gamma^{km/2}+
\gamma^{3km/2}\otimes H_m\: {\rm mod  }\: N_-\otimes N_+,\quad\\
\!\!\!\!\!\!& &\Delta(H_{-m})=H_{-m}\otimes
\gamma^{-3km/2}+\gamma^{-km/2}
\otimes  H_{-m}\: {\rm mod  }\: N_-\otimes N_+,
\quad\\
\!\!\!\!\!\!& &\Delta(q^{\pm \sq2 H_0})=
q^{\pm \sq2 H_0}\otimes   q^{\pm \sq2 H_0},\quad\\
\!\!\!\!\!\!& &\Delta(\gamma^{\pm \half})=
\gamma^{\pm\half}\otimes
\gamma^{\pm \half},\\
\!\!\!\!\!\!& &\Delta(q^{\pm d})=q^{\pm d}\otimes q^{\pm d},
\label{comult}\er\ee
where $m>0$, $n\geq 0$, and $N_\pm$ and $N_\pm^2$ are
left ${\bf Q}(q)[\gamma^\pm,  \Psi_m, \Phi_{-n}; \: m, n\in
\Z_{\geq 0}]$-modules
generated
 by $\{E^\pm_m; \:m\in  \Z\}$
and $\{E^\pm_m E^\pm_n; \:m, n\in \Z\}$  respectively
\cite{ChPr91,Jimal92}.
The second presentation of $\uq$  is the quantum
analogue of the one given in (\ref{da}) of
the $\LS$ classical algebra.

There is yet a third description of
$\uq$, which we shall call the
quantum current algebra and which is
the quantum analogue of the
$\LS$ current algebra  defined in
(\ref{ope1})-(\ref{ope4}).
This third form will again be most convenient
when dealing with free field realizations. It is
defined through the following OPE's:
\bea
\Psi(z).\Phi(w)&=&
{(z-wq^{2+k})(z-wq^{-2-k})\over (z-wq^{2-k})(z-wq^{-2+k})}
\Phi(w).\Psi(z),\label{op1} \\
\Psi(z).E^{\pm}(w)&=&
q^{\pm 2}{(z-wq^{\mp(2+k/2)})\over z-wq^{\pm (2-k/2)}}
E^\pm(w).\Psi(z),\label{op2} \\
\Phi(z).E^{\pm}(w)&=&
q^{\pm 2}{(z-wq^{\mp(2-k/2)})\over z-wq^{\pm (2+k/2)}}
E^\pm(w).\Phi(z),\label{op3}\\
E^+(z).E^-(w)&\sim& {1
\over w(\q)}\left\{{\Psi(wq^{k/2})\over z-wq^k}-
{\Phi(wq^{-k/2})\over z-wq^{-k}}\right\},\quad |z|>|wq^{\pm k}|,
\label{op4}\\
E^-(z).E^+(w)&\sim& {1
\over w(\q)}\left\{{\Psi(wq^{k/2})\over z-wq^k}-
{\Phi(wq^{-k/2})\over z-wq^{-k}}\right\},\quad |z|>|wq^{\pm k}|,
\label{op5}\\
E^{\pm}(z).E^{\pm}(w)&=&{(z q^{\pm 2}-w)\over z-w q^{\pm 2}}
E^{\pm}(w). E^{\pm}(z).
\label{op6}
\eea
Here, the quantum currents $E^\pm(z)$ are the
following generating functions of the Drinfeld generators:
\be
E^\pm(z)=\sum_{n\in\Z}E^\pm_nz^{-n-1},
\ee
and $\Psi(z)$ and $\Phi(z)$ are given
by (\ref{algebra}). The combination $(\Psi(z)-\Phi(z))/(\sq2 z(\q))$
is the quantum analogue of the classical
current $H(z)$. Note that we have used
the same symbol, $E^\pm(z)$, to denote both
the classical
and quantum currents
 but it should be clear from the context
which ones are being considered.
In order to a find a free field realization of the
$\uq$ currents $E^\pm(z)$, $\Psi(z)$ and $\Phi(z)$ one
therefore needs to resolve the relations
(\ref{op1})-(\ref{op6}) in terms of free fields, either bosonic
or fermionic.

\subsection{Bosonic realization of $\uq$ at level 2}

The bosonization of $\uq$ has recently attracted a
lot of interest. First, Frenkel and Jing derived
a bosonization   of this algebra at level
$k=1$ in terms of a single deformed
free bosonic field \cite{FrJi88}. Then,
many authors generalized this
bosonization to a generic level $k$ in terms of three
deformed free bosonic fields. These results are
referred to as  q-deformations
of the Wakimoto bosonization
(see \cite{Bou93} for a complete list of references.)
Finally, the q-deformation of the
Wakimoto bosonization of $U_q(\widehat{sl(n)})$ has been
constructed in Ref. \cite{AOSal93}, and  realizations
of the $su(n)$ finite Lie
algebra have been independently achieved in both
Refs. \cite{FlVi93} and \cite{ANOal93}.
Let us now briefly review  the  bosonization
introduced in \cite{Abaal92} (see also \cite{Bou93})
since it is the direct
quantum analogue of the classical bosonization given
in (\ref{H}). For
the case that we are interested in, i.e., $k=2$,
great simplifications occur although most features
of the general case (such as the need for three
deformed bosonic fields) are retained.
This bosonization  is given by
\cite{Abaal92}:
\bac
\!\!\!\!\Psi(z)&= &\exp\left(i\va^{1,+}(zq)-
i\va^{1,-}(zq^{-1})  \right)\\
&=&
q^{2\va^1_0}\exp\left(
2(\q)\sum_{n>0}\va^1_nz^{-n}\right),\\
\!\!\!\!\Phi(z)&= &\exp\left(i\va^{1,+}(zq^{-1})-
i\va^{1,-}(zq)\right)\\
&=&q^{-2\va^1_0}
\exp\left(-2(\q)\sum_{n<0}\va^1_nz^{-n}\right),\\
\!\!\!\!E^{\pm}(z)&=&{\exp(\pm i\va^{1,\pm}(z))
\over z(\q)}(X^{\pm}(z)-Y^{\pm}(z)),
\label{ABE}
\ea
where
\bac
X^{\pm}(z)&=&\exp\{\pm i\va^2(zq)+
{i\over \sq2}(\va^3(zq^2)-\va^3(z
))\},\\
Y^{\pm}(z)&=&\exp\{\pm i\va^2(zq^{-1})+
{i\over \sq2}(\va^3(zq^{-2})-\va^3(z))\}.
\label{XY}
\ea
The three deformed free bosonic fields
 $\va^{1,\pm}(z)$, $\va^2(z)$ and   $\va^3(z)$ are defined by
\bac
\va^{1,\pm}(z)&=&\va^1-i\va^1_0\ln{z}
+2i\sum_{n\neq 0}{q^{\mp |n|}\over [2n]}\va^1_nz^{-n},\\
\va^j(z)&=&\va^j-i\va^j_0\ln z+i\sum_{n\neq 0}
{z^{-n}\over n}\va^j_n,\qquad j=2,3.
\ea
Their modes $\{\va^j, \va^j_n,\quad j=1,2,3\}$
generate three deformed Heisenberg  algebras with
the following commutation relations:
\bac
{[\va^j_n,\va^\ell_m]}&=&(-1)^{j-1}nI_j(n)\delta^{j,\ell}
\delta_{n+m,0},\\
{[\va^j,\va^\ell_0]}&=&(-1)^{j-1}i\delta^{j,\ell}
\qquad  j,\ell=1,2,3,
\label{qhma}
\ea
where
\bac
I_1(n)&=&{[2n]^2\over 4n^2},\\
I_2(n)&=&{1\over 2}{[4n]\over [2n]},\\
I_3(n)&=&1.
\ea
The fields $\va^i(z)$ and their modes are
normalized so that in the limit $q\ra 1$ they tend to
the classical fields $\phi^i(z)$ and their modes
introduced in (\ref{cphi}) and (\ref{hma}).
It can also be checked that in this same
limit, the bosonization of the quantum currents $E^\pm(z)$ and
$(\Psi(z)-\Phi(z))/(\sq2 z(\q))$ reduces to that of the
classical currents $E^\pm(z)$ and $H(z)$ given in
(\ref{H}).

\subsection{Fermionic realization of $\uq$ at level 2}

In this section, we construct the quantum currents
$E^\pm(z)$ and ${(\Psi(zq)-\Phi(zq^{-1}))/(z(\q))}$
in terms of one
deformed free real fermionic field and one deformed
`interactive' complex fermionic field so that the quantum
current algebra (\ref{op1})-(\ref{op6}) is satisfied. Here again
we shall use the same notation as in the classical
case and denote the deformed real fermionic field by
$\psi(z)$, and the deformed complex fermionic field and
its Hermitian conjugate by $\chi(z)$ and $\chi^{\dag}(z)$ respectively.
Although  we have originally used the
quantum analogue of the boson-fermion correspondence
to derive the fermion realization of $\uq$, we will now
in analogy with the classical case, follow the opposite
path by first stating our results and  justifying them
afterwards in the next section where we elaborate on this
quantum boson-fermion correspondence. Our fermionic
realization will be given by
\bac
E^+(z)&=&\sqrt{[2]}\psi(z)\chi(z),\\
E^-(z)&=&\sqrt{[2]}\psi(z)\chi^{\dag}(z),\\
{\Psi(zq)-\Phi(zq^{-1})\over z(\q)}&=&
q:\chi(zq^2)\chi^{\dag}(z):+q^{-1}:\chi(zq^{-2})
\chi^{\dag}(z):,
\label{qfr}
\ea
where $\psi(z)$,
$\chi(z)$ and $\chi^{\dag}(z)$ satisfy the
following quantum analogue of the
Clifford current algebra:
\bea
\psi(z).\psi(w)&=&{z-w\over (z-wq^2)(z-wq^{-2})}+
:\psi(z)\psi(w):,\quad |z|>|wq^{\pm 2}|,
\label{1}\\
\chi(z).\chi^{\dag}(w)&=&{I(w)\over z-w}+
:\chi(z)\chi^{\dag}(w):,\quad |z|>|w|,
\label{2}\\
\chi^{\dag}(z).\chi(w)&=&{I(w)\over z-w}+
:\chi^{\dag}(z)\chi(w):,\quad |z|>|w|,
\label{3}\\
\chi(z).\chi(w)&=&-q^{2}{(z-wq^{-2})\over
z-wq^2}\chi(w).\chi(z),
\label{4}\\
\chi^{\dag}(z).\chi^{\dag}(w)&=&-q^{-2}
{(z-wq^2)\over z-wq^{-2}}
\chi^{\dag}(w).\chi^{\dag}(z),
\label{5}\\
I(z).\chi(w)&=&q^{2}{(z-wq^{-2})\over
z-wq^2}\chi(w).I(z),
\label{6}\\
I(z).\chi^{\dag}(w)&=&q^{-2}
{(z-wq^2)\over z-wq^{-2}}
\chi^{\dag}(w).I(z),
\label{7}
\eea
and where `the quantum analogue of the identity operator
$I(z)$' is defined by
\bac
I(z)&\equiv&\Psi(zq)-zq(\q):\chi(zq^2)\chi^{\dag}
(z):\\
&\equiv&\Phi(zq^{-1})+zq^{-1}(\q):\chi(zq^{-2})\chi^{\dag}(z):.
\label{id5}
\ea
The mode expansions of these fermionic fields and
$I(z)$ are given by
\bac
\psi(z)&=&\sum_{r\in \Z+1/2}\psi_rz^{-r-1/2},\\
\chi(z)&=&\sum_{r\in \Z+1/2}\chi_rz^{-r-1/2},\\
\chi^{\dag}(z)&=&\sum_{r\in \Z+1/2}\chi^{\dag}_rz^{-r-1/2},\\
I(z)&=&\sum_{n\in \Z}I_nz^{-n}.
\ea
The fermionic normal ordering is again the same
as in the classical case. Consequently, one can easily
show that the relation (\ref{pp}) is also valid in the
quantum case, i.e.,
\be
:\psi(z)\psi(z):=0.
\label{id4}
\ee
Using this fermionic realization we can show for instance
that relation (\ref{4}) of the
$\uq$ current algebra
is indeed satisfied (this example is the quantum
analogue of the one given in (\ref{cla})):
\bac
E^+(z).E^-(w)&=&[2]\psi(z)\chi(z).\psi(w)\chi^{\dag}(w)\\
&=&[2]\left({z-w\over (z-wq^2)(z-wq^{-2})}+:\psi(z)\psi(w):\right)
\left({I(w)\over z-w}+:\chi(z)\chi^{\dag}(w):\right)\\
&=&[2]\left({I(w)+(z-w):\chi(z)\chi^{\dag}(w):\over
(z-wq^2)(z-wq^{-2})}\right)+{\rm regular\>}\\
&=&{[2]\over w(q^2-q^{-2})}
\left({I(w)+wq(\q):\chi(wq^2)\chi^{\dag}(w):\over z-wq^2}
-{I(w)-wq^{-1}(\q):\chi(wq^{-2})\chi^{\dag}(w):\over
z-wq^{-2}}\right)+{\rm regular\>}\\
&=&{1\over w(\q)}
\left({\Psi(wq)\over z-wq^2}
-{\Phi(wq^{-1})
\over
z-wq^{-2}}\right)+{\rm regular\>}.
\ea
The relations
(\ref{id5}) and (\ref{id4}) have been used to carry the intermediate steps.
The above Clifford quantum
current
algebra (\ref{1})-(\ref{7}) amounts to the following
in terms of modes:
\bac
\{\psi_r,\psi_s\}&=&{q^{2r}+q^{-2r}\over [2]}\delta_{r+s,0},\\
\{\chi_r,\chi^{\dag}_s\}&=&I_{r+s},\\
\chi_{r+1}\chi_s+q^2\chi_s\chi_{r+1}&=&
\chi_{s+1}\chi_r+q^2\chi_r\chi_{s+1},\\
\chi^{\dag}_{r+1}\chi^{\dag}_s+q^{-2}
\chi^{\dag}_s\chi^{\dag}_{r+1}&=&
\chi^{\dag}_{s+1}\chi^{\dag}_r+
q^{-2}\chi^{\dag}_r\chi^{\dag}_{s+1},\\
I_{n+1}\chi_r-q^2\chi_rI_{n+1}&=&
-\chi_{r+1}I_n+q^2I_n\chi_{r+1},\\
I_{n+1}\chi^{\dag}_r-q^{-2}
\chi^{\dag}_rI_{n+1}&=&
-\chi^{\dag}_{r+1}I_n+
q^{-2}I_n\chi^{\dag}_{r+1},\quad n\in\Z;\quad r,s\in \Z+1/2.
\ea
It is clear from these relations, that the real quantum
fermion field
is a free field just as its classical analogue, since its modes
can be redefined so as to satisfy the classical Clifford
mode algebra (\ref{ccma}). However, since the modes of the
quantum complex
fields $\chi(z)$ and $\chi^{\dag}(z)$ cannot be redefined so
as to satisfy the classical Clifford mode algebra (\ref{ccma}), we
may therefore consider  $\chi(z)$ and $\chi^{\dag}(z)$ as `interactive'
fields. This is a major difference from the classical case.
Note that the real
fermionic field was first defined in Ref. \cite{Ber89}
for the purpose of obtaining a mixed bosonic-fermionic
realization  of $U_q(\widehat{so(2n+1)})$
at level 1.

\subsection{Quantum boson-fermion correspondence}

We will elaborate in this section on the equivalence between
the bosonic and fermionic realizations
of $\uq$ at level 2.
Contrary to the classical case
it is now the bosonization of the complex fermionic field
$\chi(z)$ that is more subtle. Let us therefore discuss
first the
bosonization of the real fermionic field $\psi(z)$.
As in the classical case,
 it is clear that
the real fermionic field $\psi(z)$  can be bosonized
in two different but equivalent forms, which we denote
by $\psi^{\pm}(z)$:
\be
\psi(z)\equiv\psi^\pm(z)={1\over z\sqrt{[2]}(\q)}
(X^{\pm}(z)-Y^{\pm}(z)),
\ee
where $X^\pm(z)$ and $Y^\pm(z)$ are defined by (\ref{XY}).
Using the basic OPE's
\bac
\va^2(z).\va^2(w)&=&{1\over 2}\ln(z-wq^2)(z-wq^{-2})+
:\va^2(z)\va^2(w):,\\
\va^3(z).\va^3(w)&=-&\ln(z-w)+
:\va^3(z)\va^3(w):,
\ea
which are easily derived from the quantum Heisenberg algebras
(\ref{qhma}) and a normal ordering similar to
the one introduced in the classical
case, we find
\bac
\psi^{\pm}(z).\psi^{\mp}(w)&=&
{z-w\over (z-w q^2)(z-wq^{-2})}+{\rm regular\> as}
\>\> z\ra wq^{\pm 2},\\
\psi^{\pm }(z).\psi^{\pm}(w)&=&
{q\over [2]}{I^{\pm}_1(w)\over z-wq^2}+
{q^{-1}\over [2]}{I^{\pm}_2(w)\over z-wq^{-2}}+
{\rm regular\> as}\>\> z\ra wq^{\pm 2},
\ea
where
\bac
I^{\pm}_1(w)&=&{q^{-2}\over w^2(q+q^{-1})(\q)^2}
(q^{-1}:X^{\pm}(wq^2)X^{\pm}(w):\\
&&-(q+q^{-1})
:Y^{\pm}(wq^2)X^{\pm}(w):+q:Y^{\pm}(wq^2)Y^{\pm}(w):),\\
I^{\pm}_2(w)&=&{q^{2}\over w^2(q+q^{-1})(\q)^2}
(q^{-1}:X^{\pm}(wq^{-2})X^{\pm}(w)\\
&&-(q+q^{-1})
:X^{\pm}(wq^{-2})Y^{\pm}(w):+q:Y^{\pm}(wq^{-2})Y^{\pm}(w):).
\ea
 It therefore follows from the above relations, that we need
to make the identifications:
\be
I^{\pm}_1(w)\equiv I^{\pm}_2(w)\equiv 1,
\label{hop}
\ee
in order to have
\be
\psi^\pm(z).\psi^\pm(w)={z-w\over (z-wq^2)(z-wq^{-2})}+
{\rm regular\> as}\>\> z\ra wq^{\pm 2}.
\ee
The identifications (\ref{hop}) can be thought of as
the quantum analogues of the
conditions (\ref{cid}) that were arrived at in the
classical situation.

We now turn to the bosonization of the complex
fermionic fields $\chi(z)$ and
$\chi^{\dag}(z)$. Looking at the
bosonization and the fermionization of $E^\pm(z)$,
we are led to set
\bac
\chi(z)&\equiv&e^{i\va^{+,1}(z)},\\
\chi^{\dag}(z)&\equiv&e^{-i\va^{-,1}(z)}.
\label{id3}
\ea
Furthermore, by comparing the relations
(\ref{2}) and (\ref{3}) respectively with
\bea
e^{i\va^+(z)}.e^{-i\va^-(w)}&=&
{:e^{i\va^+(z)-i\va^-(w)}:\over z-w}\nonumber\\
&=&{:e^{i\va^+(w)-i\va^-(w)}:\over z-w}+
{:e^{i\va^+(z)-i\va^-(w)}:-:e^{i\va^+(w)-i\va^-(w)}:
\over z-w},\\
e^{-i\va^-(z)}.e^{i\va^+(w)}&=&
{:e^{-i\va^-(z)+i\va^+(w)}:\over z-w}\nonumber\\
&=&{:e^{i\va^+(w)-i\va^-(w)}:\over z-w}+
{:e^{-i\va^-(z)+i\va^+(w)}:-:e^{i\va^+(w)-i\va^-(w)}:
\over z-w},
\eea
we get the identifications
\bea
I(w)&\equiv& :e^{i\va^+(w)-i\va^-(w)}:,
\label{id0}\\
:\chi(z)\chi^{\dag}(w):&\equiv&
{:e^{i\va^+(z)-i\va^-(w)}:-:e^{i\va^+(w)-i\va^-(w)}:
\over z-w},
\label{id1}\\
:\chi^{\dag}(z)\chi(w):&\equiv&
{:e^{-i\va^-(z)+i\va^+(w)}:-:e^{i\va^+(w)-i\va^-(w)}:
\over z-w}.
\label{id2}
\eea
Note that the terms on the right hand side of
 (\ref{id1}) and (\ref{id2})
are regular as they should since
those on the left hand side are regular by definition.
So far, we have described the quantum boson-fermion correspondence
  for the currents $E^\pm(z)$. Let us now
consider the currents $\Psi(z)$,
$\Phi(z)$ and $I(z)$.
We have given the bosonization of $\Psi(z)$ and
$\Phi(z)$ in (\ref{ABE}).
As for the fermionization, it is only
for the combination
$(\Psi(zq)-\Phi(zq^{-1}))/z(\q)$
that we have been at this point able to derive it-see
(\ref{qfr}).
The equivalence of the bosonization and the fermionization of
the latter combination is due to the identifications
(\ref{id1}) and (\ref{id2}). As for $I(z)$, this equivalence
is due to (\ref{id0}). Because of this and
in view of (\ref{id5}), we can explicitly construct all the modes
$\{\Psi_0-\Phi_0,\Psi_{n>0},\Phi_{n<0},I_{n\neq 0};
n\in\Z_{\neq 0}\}$ in terms of the complex fermionic
ones $\{\chi_r,\chi^{\dag}
_r;r\in \Z+1/2\}$. Explicitly, we have
\bac
\Psi_n&=&q^n(\q)\sum_{r\in \Z+1/2}(q^{2r}+q^{-2r})
:\chi_r\chi^{\dag}_{n-r}:,\quad n>0,\\
\Phi_n&=&-q^{-n}(\q)\sum_{r\in \Z+1/2}(q^{2r}+q^{-2r})
:\chi_r\chi^{\dag}_{n-r}:,\quad n<0,\\
\Psi_0-\Phi_0&=&\Psi_0-\Psi^{-1}_0=
(\q)\sum_{r\in \Z+1/2}(q^{2r}+q^{-2r})
:\chi_r\chi^{\dag}_{-r}:,\\
I_n&=&q^{-n}\Psi_n-(\q)\sum_{r\in \Z+1/2}q^{-2r}
:\chi_r\chi^{\dag}_{n-r}:,\quad n>0,\\
I_n&=&q^{n}\Phi_n+(\q)\sum_{r\in \Z+1/2}q^{2r}
:\chi_r\chi^{\dag}_{n-r}:,\quad n<0.
\label{moex}
\ea
Note that the first three relations of (\ref{moex}) are
the quantum analogues of the classical relations (\ref{bla}).

\section{Conclusions}
In this paper, we have used the $\uq$ quantum algebra at
level 2 as a framework to discuss the quantum analogue
of the boson-fermion correspondence. The main virtue of this
algebra for our purposes, is that it allows both boson-real fermion and
boson-complex fermion correspondences. We have shown that the real
quantum fermion is free like its classical analogue since
it can be redefined to satisfy the same anticommutation relations
as the classical ones. However,  we have stressed that the natural
complex quantum fermion which arises in this construction,
unlike its classical counterpart, is
not free, since its fusion with
its conjugate gives rise to the new field $I(z)$. Therefore,
it cannot be redefined so as to satisfy the classical
anticommutation relations.
In our fermionic realization of $\uq$ at level 2,
the modes of the currents
$E^\pm(z)$, $(\Psi(z)-\Phi(z))/z(\q)$
and $I(z)$ are all explicitly constructed in terms of the
fermionic operators. Only the zeroth mode of $I(z)$ could not
be realized in terms of the fermionic modes.
The reason for this is still
unclear and it would be interesting to elucidate this point.
Another important issue worth looking at is the realization
of the complex interactive fermion field in terms
of  free fermion fields. This would considerably
simplify the construction of the  fermionic Fock spaces
and of the fermionic vertex operators which intertwine
them. Moreover, this would provide another method
for computing important physical quantities like
the correlation functions, the form factors and the
S matrix elements of
the spin-1 XXZ Heisenberg chain
and its continuum limit.
\section*{Acknowledgements}
A.H.B. is grateful to NSERC for providing him
with a postdoctoral fellowship. The work of L.V.
is supported through funds provided by
NSERC (Canada) and FCAR (Qu\'ebec).
We wish to thank  J. Ding for
an informative discussion.
\pagebreak


\begin{thebibliography}{10}

\bibitem{Daval92}
B.~Davies, O.~Foda, M.~Jimbo, T.~Miwa, and A.~Nakayashiki.
\newblock {\em Comm. Math. Phys.}, 151:89, 1993.

\bibitem{Luk93}
S. Lukyanov.
\newblock Free Field Representation For
Massive Integrable Models.
\newblock preprint RU-93-30.

\bibitem{DiFr93}
J. Ding and I.B. Frenkel.
\newblock {\em Comm. Math. Phys.}, 156:277, 1993.

\bibitem{Hay90}
T. Hayashi.
\newblock {\em Comm. Math. Phys.}, 127:129, 1990.



\bibitem{Fre81}
I.B. Frenkel.
\newblock {\em J. Funct. Anal.}, 44:259, 1981.


\bibitem{Gin88}
P. Ginsparg.
\newblock {\it Applied Conformal Field Theory.}
\newblock  Les Houches, Session XLIX, 1988,
\newblock {\it Champs, Cordes et ph\'enom\`enes Critiques,}
\newblock Ed. E. Br\'ezin and J. Zinn-Justin.

\bibitem{GoOl86}
P. Goddard and D. Olive.
\newblock {\em Int. J. Mod. Phys.}, A1:303, 1986.

\bibitem{Bil89}
A. Bilal.
\newblock {\em Phys. Lett.}, B236:272, 1989.

\bibitem{Dri85}
V.~G. Drinfeld.
\newblock {\em Soviet Math. Doklady}, 32:254, 1985.


\bibitem{Jim85}
M.~Jimbo.
\newblock {\em Lett. Math. Phys.}, 10:63, 1985.

\bibitem{Dri86}
V.~G. Drinfeld.
\newblock Proc. ICM, Am. Math. Soc., Berkeley, CA, 1986.


\bibitem{Dri88}
V.~G. Drinfeld.
\newblock {\em Soviet Math. Doklady}, 36:212, 1988.

\bibitem{ChPr91}
V. Chari and A. Pressley.
\newblock {\em Comm. Math. Phys.}, 142:261, 1991.

\bibitem{Jimal92}
M.~Jimbo, K.~Miki, T.~Miwa, and A.~Nakayashiki.
\newblock Correlation functions of the {X}{X}{Z}
model for {${\D}<-1$}, 1992.
\newblock RIMS preprint.

\bibitem{FrJi88}
I.B. Frenkel and N. Jing.
\newblock {\em Proc. Natl. Acad. Sc.}, 85:9373, 1988.

\bibitem{Bou93}
A.H. Bougourzi.
\newblock {\em Nuc. Phys.}, B404:457, 1993.

\bibitem{AOSal93}
H.~Awata, S.~Odake, and J.~Shiraishi.
\newblock Free boson realization of $U_q(\widehat{sl_N})$, 1993.
\newblock {RIMS-924}, {YITP/K-1018} preprint.

\bibitem{FlVi93}
R. Floreanini and L. Vinet.
\newblock {\em Phys. Lett. } B315:299, 1993.


\bibitem{ANOal93}
H.~Awata, M. Noumi and S.~Odake.
\newblock Heisenberg realization of $U_q(sl(n))$ on the
flag manifold, 1993.
\newblock  {YITP/K-1016} preprint.

\bibitem{Abaal92}
A. Abada, A.H. Bougourzi and M.A. El Gradechi.
\newblock {\em Mod. Phys. Lett.},  A8:715, 1993.

\bibitem{Ber89}
D. Bernard.
\newblock {\em Lett. Math. Phys.}, 17:239, 1989.

\end{thebibliography}
\end{document}